# First-Order Mixed Integer Linear Programming

**Geoffrey J. Gordon**
Machine Learning Department
Carnegie Mellon University
Pittsburgh, PA 15213
ggordon@cs.cmu.edu

**Sue Ann Hong**
Computer Science Department
Carnegie Mellon University
Pittsburgh, PA 15213
sahong@cs.cmu.edu

**Miroslav Dudík**
Machine Learning Department
Carnegie Mellon University
Pittsburgh, PA 15213
mdudik@cs.cmu.edu

**Abstract**

Mixed integer linear programming (MILP) is a powerful representation often used to formulate decision-making problems under uncertainty. However, it lacks a natural mechanism to reason about objects, classes of objects, and relations. First-order logic (FOL), on the other hand, excels at reasoning about classes of objects, but lacks a rich representation of uncertainty. While representing propositional logic in MILP has been extensively explored, no theory exists yet for fully combining FOL with MILP. We propose a new representation, called first-order programming or FOP, which subsumes both FOL and MILP. We establish formal methods for reasoning about first order programs, including a sound and complete lifted inference procedure for integer first order programs. Since FOP can offer exponential savings in representation and proof size compared to FOL, and since representations and proofs are never significantly longer in FOP than in FOL, we anticipate that inference in FOP will be more tractable than inference in FOL for corresponding problems.

## 1 INTRODUCTION

Mixed integer linear programming has established itself as a successful formalism for decision-making under uncertainty in operation research and cooperative control, and has attracted attention more recently in AI and machine learning. For example, in OR, a common approach to solving multi-stage planning problems under uncertainty is stochastic programming with recourse (e.g., [Powell, 1996]); or, in AI, we can use MILPs for planning [Vossen et al., 1999] or to reason about uncertainty due to the actions of other agents in a nonzero-sum game [Sandholm et al., 2005]; or, in ML, we can use MILPs for MAP inference in graphical models (e.g., [Roth and Yih, 2005]); or, in cooperative control, we can use MILPs for task allocation under uncertainty [Alighanbari and How, 2005].

Many decision problems naturally contain objects, classes of objects, and relations among them. In such problems, there are many benefits to reasoning about entire classes of objects at once—so-called *lifted* reasoning. One benefit is representational: it is much simpler to state a single lifted constraint such as "all cars must follow the speed limit" than to state the constraint for each car separately. Another is computational: if we can derive a conclusion for all class members at once, then we don't need to derive it separately for each individual object. Lifted inference may incur some initial overhead, but its cost is independent of the number of objects involved, even when this number is infinite. A final benefit is statistical: if we can share parameters among members of a class, we can often reduce the number of parameters we need to estimate, increasing the level of accuracy we can attain for a given amount of data.

Unfortunately, MILPs lack an inherent mechanism to reason about classes of objects and relations. One might try to add this capability to MILPs using a "wrapper" or "compiler" such as AMPL [Fourer et al., 2002]: for example, to express a constraint that holds for all objects in a class, we can generate one copy of this constraint for every known object. However, in addition to losing the computational benefits of lifting, such a mechanism does not allow for reasoning when the number of objects is unknown or infinite, as is the case for example in entity resolution.

First-order logic, on the other hand, is designed for lifted representation and reasoning: it uses unary predicates to refer to classes of objects, higher-arity predicates to refer to relations, and quantifiers to state properties that apply to many objects or tuples of objects at once. And, unification and resolution work on quantified statements directly, allowing us to reason efficiently about entire classes of objects. Unfortunately, FOL lacks a rich representation of uncertainty: in FOL, an atom's truth value is either perfectly known or completely unknown.

To take advantage of the benefits of both languages, we introduce a new representation, *first-order programming*, which subsumes both first order logic and mixed integer linear programming. FOP has syntax and semantics similar to FOL, but includes real and integer-valued predicates, analogous to MILP variables. We establish formal methods for reasoning in FOP, including a sound and complete lifted inference procedure for the integer fragment of FOP. (Our inference procedure works for mixed-integer FOP as well, but provides only weaker guarantees in this case.) To derive our procedure, we lift the well-known Gomory cutting plane algorithm; other families of cuts for MILPs could also lead to useful inference rules, but we focus on Gomory cuts for concreteness.

We proceed as follows. We first define a syntax and semantics for the FOP language, and give examples of problems represented in FOP. We then show how to transform FOP sentences to a normal form. Working from the normal form, we describe our inference procedure and prove its soundness and completeness. Finally, we discuss extensions, outline future work, and conclude.

## 2 RELATED WORK

It is well understood that we can combine propositional logic with integer programs, or translate one to the other. For example, Hooker and Osrio [1999] give an overview of the area and a general framework called mixed logical/linear programming. To handle the more general language of FOL, we can Herbrandize a sentence and translate it to a possibly-infinite MILP (cf. Sec. 5). Several authors have used this line of reasoning to bring optimization tools and theory to bear on FOL inference [Borkar et al., 2002, Chandru and Hooker, 1999]; results include compactness lemmas analogous to our Lemma 5.3 (but specialized to FOL), as well as ways to organize the search for a finite proof of infeasibility. Some of these search methods can be thought of as lifted inference procedures for FOL. However, none of these works extend the semantics of FOL, an essential contribution of FOP.

Eaves and Rothblum [1994] provide a syntax and semantics for mathematical programs with FOL-like quantifiers and connectives, called *linear problems*. However, the quantifiers are defined over assignments of numeric values to variables, not over assignments of objects as in FOL and FOP. Therefore, linear problems cannot reason about classes of objects, and do not generalize FOL. (In fact, Eaves and Rothblum show that linear problems are decidable in finite time, which implies that the language of linear problems is strictly less expressive than FOL.)

Relational probabilistic languages (RPLs), such as plate models, probabilistic entity-relationship models [Heckerman et al., 2007], Markov logic networks [Richardson and Domingos, 2006], IBAL [Pfeffer, 2001], ICL [Poole, 2008], and BLOG [Milch et al., 2005], combine probabilistic and logical representations, and allow us to reason about uncertain statements about classes of objects. Hence, the goals of RPLs are in many ways similar to the goal of FOP. However, the focus of FOP is different from the focus of RPLs: while FOP is designed to capture uncertainty through *nondeterministic choice* and *optimization* (as in FOL and MILPs), RPLs are designed to capture uncertainty through *marginalization* and *factorization* (as in graphical models). Since FOP is Turing-complete, as are many RPLs, it is technically possible to translate any program from either type of representation to the other. But, for any given reasoning problem, we expect that the difference in representation styles means that either sort of language may lead to a much more natural and compact representation than the other.

Additionally, some RPLs do not provide semantics for unknown or infinite numbers of objects. (Exceptions include BLOG [Milch et al., 2005], which allows unknown numbers of objects; ICL [Poole, 2008], which allows unknown or infinite numbers of objects; and the work of Singla and Domingos [2007], which extends Markov logic to infinite domains, but makes strong assumptions in order to guarantee well-defined probability distributions.) Lifted inference algorithms have been proposed for RPLs, but many work only when the number of objects is finite and known; e.g., the work of Braz et al. [2005] on lifted variable elimination requires a known set of objects, and lifted belief propagation [Singla and Domingos, 2008] was described in terms of finite MLNs, leaving infinite domains explicitly for future work. In contrast, full lifted inference for first-order logic is well understood (see, e.g., Russell and Norvig [2003]), and our inference procedure for FOP parallels the well-known resolution procedure for FOL.

## 3 DEFINITION OF FOP

### 3.1 SYNTAX

The syntax of FOP parallels that of FOL. *Terms* are expressions representing *objects*; *formulas* are those representing *values*. However, where there are only two values in FOL (true and false), values in FOP are bounded reals or integers.[1] Just as FOL has the constant literals $T$ and $F$, FOP has *scalars*, which are literals with predefined real values. Like FOL, FOP contains *functions* that map objects to objects, and *predicates* that map objects to numeric values. An *atom* is a predicate applied to a tuple of objects. A *literal* is a scalar multiple of an atom.

A FOP formula is a string of literals combined using operators and quantifiers, as recursively defined in Fig. 1. Quantifiers for FOP are *supremum* and *infimum*, analoguous to

---

[1] Boundedness is crucial to the completeness of our inference procedure; however, there is no limit on how large the bounds can be, so this requirement presents little practical concern.

∃ and ∀ in FOL repectively. A variable with a matching quantifier is *bound*, while one without a matching quantifier is *unbound* or *free*. Binary addition, subtraction, min, and max are allowed between formulas. *Scalar multiplication*, i.e., multiplication of formulas by scalars, is allowed, but multiplication of two general formulas is not, in order to maintain linearity. Operators have the usual precedence order: scalar multiplication (including negation), binary operators $\{+, -\}, \wedge, \vee$, then quantifiers. As usual, a *ground term* is a term containing no unbound variables, and a *sentence* is a formula containing no unbound variables.

FOP has two constructs analogous to FOL clauses, a *sum-clause* (a sum of literals), and a *max-clause* (a maximum of literals). We use *clause* to refer to either one, when clear from context. A *superclause* is a maximum of sum-clauses; FOL has no corresponding construct.

### 3.2 Relationship to FOL and MILP

We give a formal semantics for FOP below, in Sec. 3.3. For now, we informally say that a *model* in FOP is analogous to a model in FOL: a list of the objects in our world and a table of values for each function and predicate. Given a FOP sentence and model, we can evaluate the sentence by looking up its objects, functions, and predicates in the model, then combining the results using the standard definitions of operators like ∨ and +. A sentence's *value* is the number to which it evaluates under the best (maximizing) model.

With this definition, the relationship between FOL and FOP is simple: we can translate any FOL sentence to FOP, preserving its value under all models, and therefore preserving its satisfiability (its maximal value under any model). Table 1 shows two possible translations. While Trans. A is more direct, Trans. B allows us to identify our lifted Gomory cuts (Sec. 5) as a direct generalization of FOL resolution (see Gordon et al. [2009]).

As an example, consider the following sentence in FOL:

$(\text{bird}(x) \Rightarrow \text{flies}(x)) \quad \wedge \quad (\text{eagle}(y) \Rightarrow \text{bird}(y)) \quad \wedge$
$(\text{eagle}(z) \Rightarrow \text{eagle}(\text{father}(z))) \quad \wedge \quad \text{eagle}(\text{Stanley})$

which can be written in FOP (using Trans. B and some further manipulations) as follows:

$(\text{flies}(x) - \text{bird}(x)) \quad \wedge \quad (\text{bird}(y) - \text{eagle}(y)) \quad \wedge$
$(\text{eagle}(\text{father}(z)) - \text{eagle}(z)) \quad \wedge \quad (\text{eagle}(\text{Stanley}) - 1)$

In the above sentences, we have used the convention that free variables are implicitly quantified, using ∀ (in FOL) or ⋀ (in FOP)—e.g., the first conjunct in the FOP sentence can be read as $\bigwedge x.\ (\text{flies}(x) - \text{bird}(x))$. The FOP sentence is incomplete unless we specify the ranges of the predicates (flies, bird, and eagle); in this case, all take values in $\{0, 1\}$.

It is easy to see that every satisfying model for the FOL sentence corresponds to a model which makes the FOP sentence nonnegative. In FOL, we may ask whether, given our

**Figure 1** Definition of FOP Terms and Formulas

**Operators, Quantifiers**:
1. Negation −, scalar multiplication ∗
2. Binary addition +, subtraction −, max ∨, min ∧
3. Quantifiers: inf ⋀, sup ⋁

**Terms**: A term represents an object.
1. A *variable* representing objects
2. A *function* applied to 0 or more objects: fun$(t_1,...,t_n)$

**Atoms**: An atom evaluates to a number.
1. A *scalar*, a real-valued constant
2. A *predicate* applied to 0 or more terms: pred$(t_1,...,t_n)$. Each predicate is annotated with its *range*, a bounded interval of ℝ or ℤ.

**Formulas (Statements)**: A *formula* is an expression that evaluates to a number, and is defined recursively:
1. Any atom is a formula.
2. If $f$ is a formula and $c$ is a scalar, then the following are formulas: $-f, f * c, c * f$
3. If $f, g$ are formulas, then so are the following: $f + g$, $f - g$, $f \vee g$, $f \wedge g$.
4. If $f$ is a formula and $x$ is a variable, then the following are formulas: $\bigwedge x.\ f, \bigvee x.\ f$

---

Table 1: Two possible translations of FOL to FOP

| FOL | Trans. A | Trans. B |
|---|---|---|
| $T$ | 1 | 1 |
| $F$ | $-1$ | 0 |
| $P \wedge Q$ | $P \wedge Q$ | $P \wedge Q$ |
| $P \vee Q$ | $P \vee Q$ | $1 \wedge (P + Q)$ |
| $\neg P$ | $-P$ | $1 - P$ |
| $\forall x.\ P$ | $\bigwedge x.\ P$ | $\bigwedge x.\ P$ |
| $\exists x.\ P$ | $\bigvee x.\ P$ | $\bigvee x.\ P$ |

assumptions, we can conclude flies(father(Stanley)), and use resolution to show that Stanley's father flies. In FOP, we can use our lifted Gomory cut procedure to show that our assumptions imply flies(father(Stanley)) ≥ 1.

Table 1 shows that it is never necessary to expand representation size when going from FOL to FOP. In fact, the reverse is true: we can sometimes achieve an exponentially smaller representation. For example, to describe the property "$p(x)$ for at least $k$ of a given $n$ objects" in FOL, we need an exponentially-long formula: one such formula is the disjunction of $\binom{n}{k}$ clauses, each of which tests $p(x)$ for one subset of $k$ objects (e.g., $p(x_1) \wedge \ldots \wedge p(x_k)$). In FOP, on the other hand, we can describe the same property with a formula of length linear in $n$: if $p(x) \in \{0, 1\}$, then $p(x_1) + p(x_2) + \ldots + p(x_n) - k$ has value $\geq 0$ exactly when $p(x) = 1$ for at least $k$ of $x_1, \ldots, x_n$.

Similarly, we can translate MILPs to FOP. To determine the feasibility of a MILP, we translate each of its constraints into a sum-clause, and conjoin all of these clauses (using $\wedge$). For example, consider the MILP schema

$$\begin{aligned} \max x_1 \text{ s.t.} & & & \\ x_i & \geq & 2x_{S(i)} & \forall i \in I \\ x_i & \in & \{0, 1, \ldots, 8\} & \forall i \in I \end{aligned} \quad (3.1)$$

Here $I$ is an unspecified finite index set and $S : I \to I$ is an unspecified function. We instantiate the schema to get an ordinary MILP by fixing $I$ and $S$; for example, if we set $I = \{1, 2, 3, 4\}$ and $S(i) = \min(i + 1, 4)$, an optimal solution is $(x_1, x_2, x_3, x_4) = (8, 4, 2, 0)$.

We can translate the constraints from (3.1) to the FOP sentence $(x(i) - 2 * x(S(i)))$, where $x$ is a one-argument predicate with range $\{0, 1, \ldots, 8\}$. We can also translate the objective [Gordon et al., 2009], or simply test whether there is a solution of value $c$ by adding the clause $(x(1) - c)$.

Just like the MILP schema, our FOP sentence doesn't specify the function $S$ or the index set $I$. As we did above, we can specify $S$ and $I$ before inference. But, we have another choice: we can leave $S$ and $I$ unspecified or only partly specified, and ask our FOP inference algorithm to *discover* values for $S$ and $I$. We can also ask our FOP inference algorithm to tell us facts about $x$, $S$, and $I$ which must be true in *any* solution of (3.1); see Sec. 5.4 for an example.

Just as for FOL, translating a MILP or MILP schema to FOP cannot increase representation size. On the other hand, it can be impossible to translate a FOP sentence to a MILP or MILP schema, since MILPs and MILP schemas are not Turing-complete, while FOP includes FOL.

As indicated by the above example, FOP variables and MILP variables do *not* correspond to one another. A MILP variable corresponds to a FOP ground atom. A FOP variable, on the other hand, can be thought of as an index into a family of related MILP variables, since a FOP expression like $x(i) - 2 * x(S(i))$ expresses a whole family of constraints $x_i \geq 2x_{S(i)}$, one for each object $i$ could refer to.

### 3.3 SEMANTICS

A model $M$ is a tuple $(O, F, P)$, where $O$ is a list of objects, $F$ is a table of function values, and $P$ is a table of predicate values. A valuation $I$ (under $M$) is a mapping of syntactic variables to model objects. We write $\mathcal{M}$ for the set of possible models, and $\mathcal{I}(M)$ for the set of possible valuations under $M$. In the context of a model, each expression of FOP evaluates to a *function* mapping a valuation to a value (a model object or a number). (We need to use a function since in the inner stages of recursive evaluation we do not yet know how free variables will be quantified.) A ground expression then evaluates to a constant function, mapping every valuation to the same object or real number.

In particular, in the context of a model $M$, a sentence $S$ evaluates to a constant function, mapping every valuation to the same real number. We will abuse notation slightly and write $\text{value}(S, M)$ to mean this real number. Finally, we define $\text{value}(S)$ to be $\sup_{M \in \mathcal{M}} \text{value}(S, M)$.

We evaluate an expression under a model compositionally, based on its outermost syntactic operation:

1. If the expression is a variable (say $x$), we return the function which looks that variable up in a valuation: $\lambda I. \, I(x)$. (Here and below, $\lambda I. \, [\ldots]$ refers to a function taking a single argument $I \in \mathcal{I}(M)$, a valuation.)

2. Scalar constants evaluate to constant functions: e.g., the formula 3 evaluates to $\lambda I. \, 3$.

3. If the outermost operation is an arithmetic operator, we compose by ordinary arithmetic: e.g, for $A + B$, if $A$ evaluates to a function $a : \mathcal{I}(M) \to O$ and $B$ evaluates to a function $b : \mathcal{I}(M) \to O$, then $A + B$ evaluates to $\lambda I. \, a(I) + b(I)$. Negation, scalar multiplication, and binary $-$, $\wedge$, and $\vee$ are analogous.

4. If the outermost operation is application of a function $f$, we first evaluate each argument expression, getting argument functions $a_1, \ldots, a_k$, each in $\mathcal{I}(M) \to O$. We then return $\lambda I. \, F(f, a_1(I), \ldots, a_k(I))$: given a valuation $I \in \mathcal{I}(M)$, we find the value of each argument under $I$, then look up $f$ and $a_1(I), \ldots, a_k(I)$ in our function table $F$. (Object constants are just zero-argument functions, and so evaluate by simple lookup in $F$, with no recursive evaluation of arguments.)

5. Similarly, for application of a predicate $p$, we we first evaluate each argument expression, getting functions $a_1, \ldots, a_k$, each in $\mathcal{I}(M) \to O$. We then return $\lambda I. \, P(p, a_1(I), \ldots, a_k(I))$, which looks up $p$ and its arguments in the predicate table $P$. (Real or integer-valued variables are just zero-argument predicates, and therefore evaluate by simple lookup in $P$.)

6. The quantifier expression $\bigwedge x. \, Q$ works by taking an infimum over model objects: write $I[x \to o]$ for the valuation which is the same as $I$ except that it maps the variable $x$ to the object $o \in O$. Then, if $Q$ evaluates to $q : \mathcal{I}(M) \to \mathbb{R}$, then $\bigwedge x. \, Q$ evaluates to $\lambda I. \, \inf_{o \in O} q(I[x \to o])$. The quantifier expression $\bigvee x. \, Q$ is similar, substituting sup for inf.

**Variable Substitution.** We define a *variable substitution* $V$ as a mapping from variables to terms, for example $\{x \to John, y \to uncle(z)\}$. We write application of $V$ to an expression $S$ as $S/V$, with the usual meaning that each free variable in $S$ appearing in the domain of $V$ gets replaced by the corresponding term. (Bound variables are unaffected.) All variables appearing in the RHS of a substitution must be distinct from all variables on the LHS. Note that the process is purely *syntactic*: the result depends only on $V$ and on whether a variable is syntactically bound in $S$.

Given an expression $S$, we can also define a *variable substitution for $S$* as a variable substitution whose range contains only the object constants and functions mentioned in $S$, along with the special constant nil if $S$ contains no other object constants. With this definition, $V$ will not introduce new functions or constants into $S/V$ (except possibly nil). Finally, given two substitutions $V$ and $W$, we write composition as $V/W$, so that $S/(V/W) = (S/V)/W$.

## 4 NORMAL FORMS

To facilitate analysis, we translate sentences into one of two normal forms: *min-normal form* or *reduced normal form*. Min-normal form preserves the value of a sentence in all models, while reduced normal form preserves only the sign of the value. In both forms, we move negation and scalar multiplication inward, eliminate $\bigvee$ quantifiers, and rearrange binary operators to apply in the order $\wedge$, $\vee$, $+$ (outermost to innermost). In reduced normal form, we also eliminate $\vee$. Figs. 2 and 3 give translation procedures.

Our translation to min-normal form is analogous to the well-known *conjunctive normal form* translation in FOL. We could also define a symmetric translation to a *max-normal* form, isomorphic to the min-normal form of $-S$; as we will see, though, min-normal form lends itself to proofs of upper bounds on the value of a sentence, and therefore to proofs by contradiction.

Reduced normal form is the final form used by our inference procedure. It highlights the relationship of FOP to MILP: as mentioned above, conjoining sum-clauses with $\wedge$ (as we do in reduced normal form) is analogous to intersecting linear constraints in a MILP. FOP as a whole assigns no special role to 0, but reduced normal form is designed for comparing a sentence's value to 0. We could easily define an analogous normal form for comparing against any other threshold $k$, or simply translate $S - k$ instead of $S$ to reduced normal form.

In more detail, for min-normal form (Fig. 2), after step 1, each negation and scalar multiplication becomes part of a literal. Step 2 ensures that we do not quantify over the same variable more than once, in preparation for steps 3–4. Step 3 eliminates $\bigvee$ by using the fact that value$(S)$ contains a supremum over models. Step 4 is purely syntactic, since quantifier ordering no longer matters after Steps 2–3. Step 5 finally transforms the formula to normal form, using either distributive laws (such as $(A \wedge B) \vee C = (A \vee C) \wedge (B \vee C)$) or the more-efficient Tseitin transformation. We show [Gordon et al., 2009] that the Tseitin transformation leads to at most a constant-factor growth in the length of our formula (compared to potentially-exponential growth for the simpler distributive procedure). Hence, the min-normal-form translation (with the Tseitin procedure) maintains the compactness of FOP.

**Figure 2** Transformation to Min-Normal Form

Given a FOP sentence,

1. Move negation and scalar multiplication inward as far as possible, by distributing positive scalar multiplication and using rules analogous to De Morgan's laws:
   $-\bigwedge x.S = \bigvee x.(-S)$, $-(A \wedge B) = (-A) \vee (-B)$,
   $-\bigvee x.S = \bigwedge x.(-S)$, $-(A \vee B) = (-A) \wedge (-B)$.
2. *Standardize apart* quantified variables to unique names.
3. "Skolemize": replace statements $\bigvee x. S$ with $S/\{x \to x(L)\}$, where $L$ is the list of the names of enclosing min-quantified variables. E.g., $\bigwedge y. \bigwedge z. \bigvee x. P(x)$ becomes $\bigwedge y. \bigwedge z. P(x(y, z))$.
4. Remove all $\bigwedge$ quantifiers; by convention, all free variables will be implicitly $\bigwedge$-quantified.[2]
5. Ensure that binary operators are applied only in the order $\wedge$, $\vee$, $+$, using either distributive laws or a procedure analogous to the Tseitin transformation for propositional logic [Gordon et al., 2009].

**Figure 3** Transformation to Reduced Normal Form

Given a sentence $S = C_1 \wedge \ldots \wedge C_m$ in min-normal form, replace each superclause $C_i = (L_{i1} \vee \ldots \vee L_{in})$ with the conjunction of sum-clauses $L'_{i1} \wedge \ldots \wedge L'_{in} \wedge R_i$, where

1. $L'_{ij} = L_{ij} + B_{ij} * (1 - z_{ij}(\ldots))$, where $B_{ij}$ is a sufficiently large[3] scalar and $z_{ij}(\ldots)$ is a new 0-1 predicate whose arguments are the free variables in $C_i$
2. $R_i = z_{i1}(\ldots) + \ldots + z_{in}(\ldots) - \frac{1}{2}$.

In reduced normal form (Fig. 3) we further replace each superclause with a conjunction of sum-clauses. To do so, we introduce new variables $z_{ij}(x, y)$ to indicate which sum-clause $L_{ij}$ gives the maximum value to each original superclause $C_i$, and enforce $L_{ij}$ iff its corresponding indicator is 1. The additional sum-clause $R_i$ ensures that for every $i$ and for every setting of the free variables in $C_i$ there exists $j$ s.t. $z_{ij}(\ldots) = 1$, so that we always enforce at least one sum-clause out of each disjunction.

## 5 INFERENCE

Given a sentence in reduced normal form, the most basic reasoning question is to determine whether it is *feasible*, that is, whether its value is at least zero. Using our feasibility test as a primitive, we can place bounds on the value of a general FOP sentence by binary search. Below, we also generalize the FOL concept of *entailment*, and use feasibility to test whether one FOP sentence is a logical conse-quence of another. We will say that *inference* is the problem of generating entailed consequences of a given sentence.

---

[2]We extend our semantics in the natural way: value$(S)$, where $S$ is in a normal form, fills in quantifiers using this convention.

[3]It is enough to pick $B_{ij} \geq \text{range}(L_{ij})$, but it is easier to pick $B_{ij} \geq \sum_{P \in \mathcal{P}_{ij}} \text{range}(P) \, \text{coeff}(P, L_{ij}) \geq \text{range}(L_{ij})$. Here $\mathcal{P}_{ij}$ is the set of predicates mentioned in $L_{ij}$, range$(F)$ is the difference between the largest and smallest possible values of $F$, and coeff$(P, L_{ij})$ is the sum of absolute values of scalar multipliers in the literals based on $P$ in $L_{ij}$.

Since feasibility in FOP is analogous to satisfiability in FOL, we could try to design an inference procedure for FOP by reducing FOP to FOL. However, a good reduction is not immediately obvious; and in fact, we believe reducing to FOL would lose valuable structure in many sentences, making proofs longer and more difficult to find and interpret. Instead, we seek a procedure which works directly on the FOP representation.

In the following subsections, we first define entailment, soundness, and completeness. We then present a sound and complete, but non-lifted and therefore intractable, inference procedure. Finally, we give a lifted inference procedure, and show its soundness and completeness by comparing it to the first procedure.

### 5.1 ENTAILMENT

In this subsection, we focus on the integer fragment of FOP; we discuss entailment for mixed-integer FOP in Sec. 6.

In FOL, sentence $S$ entails sentence $S'$ (written $S \models S'$) iff $S'$ is true in every model which satisfies $S$. To extend this definition to FOP, we replace satisfiability by feasibility:

**Definition 5.1** (FOP entailment). $S \models S'$ iff, for all models $M$, $[\text{value}(S, M) \geq 0] \Rightarrow [\text{value}(S', M) \geq 0]$.

Using the translations from Table 1, together with an appropriate threshold (0 for Trans. A, $\frac{1}{2}$ for Trans. B), it should be clear that FOP entailment generalizes FOL entailment.

In FOL, we can reduce entailment checking to satisfiability checking: $S \models S'$ iff $S \wedge \neg S'$ is satisfiable. (This proof strategy is called *refutation*.) Similarly, in integer FOP, we can reduce entailment to feasibility checking. To do so, we need the following lemmas, whose proofs can be found in Gordon et al. [2009].

**Lemma 5.1.** *Given a sentence $S$ of integer FOP, we can efficiently (in low-order polynomial time in the length of $S$) compute an $\epsilon(S) > 0$ such that, for all models $M$, $\text{value}(S, M)$ is an integer multiple of $\epsilon(S)$.*

**Lemma 5.2.** *Let $S$ and $S'$ be sentences of integer FOP, and let $\epsilon = \epsilon(S')$. Then $S \models S'$ iff $S \wedge (-\frac{\epsilon}{2} - S')$ is infeasible.*

So, to search for a refutation proof of $S \models S'$ in FOP, we pass $S \wedge (-\frac{\epsilon(S')}{2} - S')$ to a feasibility checker.

We say that an inference procedure is *sound* if it generates *only* entailed sentences, and *complete* if it can generate *all* entailed sentences. Since refutation is based on feasibility checking, we say that a feasibility checker is *sound* if its reports of infeasibility are always correct, and *complete* if it can discover the infeasibility of all truly-infeasible sentences.

Assuming that we have access to a practical, sound, and complete feasibility checker for FOP, lemma 5.1 shows that refutation inference is practical, and lemma 5.2 shows that refutation inference is sound and complete. So, we turn next to the problem of feasibility checking.

### 5.2 FEASIBILITY

To try to prove $\text{value}(S) < 0$ for a mixed-integer sentence $S$ in reduced normal form, we can use the following *Naive Inference* procedure. We give more detail on each step of this procedure below.

1. *Propositionalize* (or *Herbrandize*) $S$.
2. For each *finite subproblem* of the Herbrandization:
    (a) Phrase the subproblem as a mixed-integer linear feasibility problem and pass it to a MILP solver.
    (b) If the MILP is infeasible, terminate and declare that $S$ is infeasible; otherwise, continue.

Naive Inference is not practical: we must generate and test all finite subproblems of the propositionalized $S$. So, in Sec. 5.3, we prove a "lifting lemma", showing that we can avoid propositionalization, and use fully first-order reasoning instead. But first, we analyze Naive Inference.

Our propositionalization procedure is analogous to that in FOL: we build a *Herbrand universe* $H$ containing all ground terms which could possibly be relevant to the value of $S$, and substitute them into $S$ in all possible ways. For example, the sentence $\text{height}(\text{father}(x))$ becomes

$$\text{height}(\text{father}(\text{nil})) \wedge \text{height}(\text{father}(\text{father}(\text{nil}))) \wedge \ldots$$

The result will be a "propositional program" $S(H)$, with no free variables, countably infinitely many distinct atoms, and countably infinitely many sum-clauses, such that $\text{value}(S(H)) = \text{value}(S)$. A more detailed description of the procedure can be found in Gordon et al. [2009].

Finding $\text{value}(S(H))$ is an optimization problem, $\sup_M \text{value}(S(H), M)$. It is sufficient to optimize over *Herbrand models*, i.e., models in which the set of objects is $H$, and in which the function tables implement the expected semantics (e.g., applying the function father to the object nil yields the object father(nil)). So, the only variable part of $M$ is the assignment of (real or integer) values to syntactically-distinct atoms. Write $D$ for the Cartesian product of the domains of all distinct atoms in $S(H)$, and write $M(d)$ for a model which implements an assignment $d \in D$. We then have $\text{value}(S(H)) = \sup_{d \in D} \text{value}(S(H), M(d))$.

Given $S(H)$, we formulate a subproblem by selecting a subset of sum-clauses: if $S(H) = G_1 \wedge G_2 \wedge \ldots$, where the $G_i$ are ground clauses, and if $J$ is a subset of the natural numbers, then subproblem $S_J$ is the conjunction (using $\wedge$) of all clauses $G_j$ for $j \in J$. For any $d \in D$,

$$\text{value}(S(H), M(d)) \leq \text{value}(S_J, M(d))$$

as minimizing over fewer clauses only increases the value.

Write $D_J$ for the subspace of $D$ corresponding to the atoms mentioned in $S_J$, and extend $M(\cdot)$ to apply to $d \in D_J$ by assigning predicate values not mentioned in $S_J$ arbitrarily. Since each $G_j$ contains finitely many atoms, $D_J$ has finitely many dimensions—that is, a subproblem has finitely many variables. Also note that $D_J$ can be described by finitely many linear constraints (upper and lower bounds on each dimension). Since

$$\text{value}(S_J, M(d)) \geq 0 \Leftrightarrow \forall j \in J.\ \text{value}(G_j, M(d)) \geq 0$$

and since each constraint $\text{value}(G_j, M(d)) \geq 0$ is linear in $d$, finding $d \in D_J$ such that $\text{value}(S_J, M(d)) \geq 0$ is a MILP.

Now that we have specified the Naive Inference procedure, we can show that it is sound and complete:

**Lemma 5.3** (completeness). *If* $\text{value}(S) < 0$, *Naive Inference will eventually find a subproblem* $S_J$ *s.t.* $\text{value}(S_J) < 0$ *(and so declare that $S$ is infeasible)*.

*Proof.* By the Tychonoff Product Theorem, $D$ is a compact set under the product topology, since it is the product of bounded subsets of the reals or integers (each of which is compact). Write $D_{[G_j < 0]}$ for the subset of $D$ containing exactly those assignments $d$ for which $\text{value}(G_j, M(d)) < 0$. Crucially, $D_{[G_j < 0]}$ is an *open* subset: $\text{value}(G_j, M(d))$ is a linear function of $d$, and a strict linear inequality on finitely many variables defines an open set.

If indeed $\text{value}(S) < 0$, then the open sets $D_{[G_j < 0]}$ for $j = 1, 2, \ldots$ form a cover of $D$: by the definition of $\text{value}(S)$, for any assignment $d \in D$ there exists a clause $G_j$ s.t. $\text{value}(G_j, M(d)) < 0$. Since $D$ is compact, this open cover must have a finite subcover. Let $J$ be the set of indices of any such subcover; then consider the subproblem $\text{value}(S_J)$. Since $D_{[G_j < 0]}$ for $j \in J$ is also a cover of $D$, we have $\text{value}(S_J) < 0$, so we can use this subproblem to prove $\text{value}(S) < 0$ as claimed. Since Naive Inference iterates through all possible subproblems, it will eventually find $S_J$ and declare that $\text{value}(S) < 0$. □

**Lemma 5.4** (Soundness). *If* $\text{value}(S) \geq 0$, *Naive Inference will never declare that* $\text{value}(S) < 0$.

*Proof.* Naive Inference will only declare that $\text{value}(S) < 0$ if finds an infeasible subproblem. An infeasible subproblem $S_J$ must satisfy $\text{value}(S_J) < 0$, which is not possible since $\text{value}(S_J) \geq \text{value}(S) \geq 0$. □

Combined with the proofs of correctness for our construction of the feasibility problem and our propositionalization process [Gordon et al., 2009], Lemmas 5.1–5.4 imply:

**Theorem 5.5.** *Naive Inference with entailment by refutation is a sound and complete inference method for the integer fragment of FOP.*

**Gomory cuts.** Naive Inference uses a MILP solver to check the feasbility of each subproblem. For our purposes, a particularly convenient solver is based on *Gomory cuts* for integer linear programs [Gomory, 1958], along with their generalization to mixed-integer programs. Since we will also use a lifted version of Gomory cuts for our lifted inference procedure below, for clarity, we refer to the non-lifted cuts as *propositional Gomory cuts*.

The details of Gomory cuts are available in standard texts on optimization [Wolsey and Nemhauser, 1988]; here, we only require the following properties. First, Gomory cuts work on MILPs in equality form, that is, a set of $m$ linear equality constraints on $n > m$ nonnegative variables. Second, given a linear combination of the constraints, we get a unique Gomory cut, represented as a new linear inequality on the $n$ variables.[4,5] Third, this cut is *valid*, i.e., does not remove any (mixed) integer points from the feasible region. Finally, any valid inequality can be derived via a finite sequence of Gomory cuts, in which later cuts work from the expanded set of constraints containing both the original constraints and all previous cuts.

These properties together ensure that an inference procedure based on Gomory cuts terminates in finite time: generate all tableau-based Gomory cuts by breadth-first search, add them one at a time to our set of constraints, and check for mixed-integral basic feasible solutions after each one. Eventually, either we will find a feasible point which satisfies all integrality constraints, or we will generate a valid inequality which cuts away all (real as well as integer) points from our feasible region. In the latter case, the sequence of cuts constitutes a proof of infeasibility of the subproblem, and therefore of the original sentence $S$.

### 5.3 LIFTED INFERENCE

While Naive Inference is sound and complete, it is impractical: it requires us to generate and test all finite subproblems of the Herbrandization of $S$ until we find an infeasible one, and for each subproblem it requires us to generate and test all Gomory cuts until we prove feasibility or infeasibility. Instead, we would prefer an inference procedure more like the highly-successful resolution rule for first-order logic: since resolution is a lifted inference rule, a resolution proof in FOL can ignore irrelevant clauses, and never needs to blow up the size of its assumptions by Herbrandizing them.

In order to trust our lifted inference rule, we must demonstrate that it can form the basis of a sound and complete inference procedure (as resolution does for FOL); else, we might gain more-efficient inference, but lose the ability to prove some true statements. To do so, we will provide a *lifting lemma*: given a propositional proof (such as the one found by Naive Inference using propositional Gomory cuts), the lifting lemma will show that we can duplicate each step of the proof using our lifted inference rule. So, we can conclude that we do not lose the ability to prove any true statements when we move to lifted inference, and also that the lifted proofs are no longer than the original propositional proofs. Combined with the observations that the lifted inference rule allows us to consider large numbers of propositional proofs at once, and that lifting can significantly shorten proofs, this fact suggests that lifted inference will make it much more practical to search for proofs in FOP.

---

[4]To handle inequalities, we can turn them to equalities using slack variables. We can eliminate the slack variables once we find the cut: if our original constraint is $Ax + b = s \geq 0$, with slack variables $s$, we can substitute $Ax + b$ for $s$ in the cut.

[5]The linear combination is usually chosen as a row of a simplex tableau; this restriction can help focus search, but is not necessary.

We start by describing our inference rule, which we call *lifted Gomory cuts*. This rule is a generalization of FOL's resolution rule, in which we combine two clauses by eliminating a *resolvent* literal which appears positively in one and negatively in the other. (See Gordon et al. [2009] for a proof.) However, while resolution only combines pairs of clauses, our lifted Gomory cut rule can work from any number of clauses at once. Just as resolution provers *unify* expressions across clauses to find resolvents, we can use variable substitutions to simplify our clause set before cutting. And, just as a resolution proof ends by deriving the empty clause, a proof using lifted Gomory cuts ends when we can use simple linear algebra to prove that our sentence's feasible region is empty.

Each lifted Gomory cut tries to prove $\text{value}(S) < 0$ using a subset of the sum-clauses of $S$. Just as we did for Naive Inference, we build a MILP representing the feasibility of our chosen subset of sum-clauses, and perform a Gomory cut in this MILP. However, in contrast to Naive Inference, lifted cuts can work from non-ground clauses; in this case, the resulting cut holds for many subproblems simultaneously (corresponding to many ways to substitute ground terms into our clauses). We represent the cuts for all of these subproblems as a new, lifted sum-clause which has value $< 0$ in model $M$ only if $\text{value}(S, M) < 0$. Because of this fact, conjoining this new clause to $S$ does not change whether $\text{value}(S) < 0$.

To build a lifted Gomory cut, we start from a sentence $S$ of FOP in reduced normal form. We then perform the following steps:

1. Select sum-clauses $L_1, L_2, \ldots L_p$ from $S$, along with a variable substitution $U$. We may pick the same clause twice, in which case we standardize the copies apart. As a special case, to reason about predicate bounds, we allow some $L_i$ to be *implicit clauses* of the form $p(x, y, \ldots) - l$ or $u - p(x, y, \ldots)$, where $[l, u]$ is the declared range for $p$, and $x, y, \ldots$ are standardized apart. Let $L'_i = L_i/U$ for each $i$.

2. Introduce a new, temporary MILP variable (zero-argument predicate) for each *textually distinct* atom in the clauses $L'_i$, and substitute the variables into the $L'_i$ to get *propositional* sum-clauses $L''_i$. Write $x$ for the vector of temporary variables.

3. Convert $L''_1 \wedge L''_2 \wedge \ldots$ to a mixed integer feasibility problem $P$ as described for subproblems in Naive Inference. This step introduces slack variables $s_i \geq 0$.

4. Pick a linear combination of constraints, and use the (propositional) Gomory method to derive a new, valid inequality $C \geq 0$ for $P$, where $C$ contains $x$ and $s_i$.

5. Eliminate the slack variables, as described previously for Naive Inference, leaving an inequality $C' \geq 0$ written in terms of $x$ only. For convenience, we allow *weakening* the cut by increasing the coefficient of each $s_i$ before eliminating it. (Weakening is never necessary, but can help interpretation.)

6. Replace each temporary variable in $C'$ with its corresponding atom, yielding a *lifted* inequality $L \geq 0$.

7. Ensure no object variable in $L$ conflicts with those in $S$ by standardizing apart $L$. Call the result $L'$. Set $S' = S \wedge L'$, and use $S'$ to continue generating lifted Gomory cuts.

Note that, while there are well-studied heuristics for FOL which choose a subset of clauses and a variable substitution intelligently, so that unification and resolution are likely to result in refutation quickly, the corresponding search control problem for FOP is a subject of future research.

Detailed proof sketches of the following two lemmas can be found in Gordon et al. [2009].

**Lemma 5.6** (Soundness). *Lifted Gomory cuts, with proof by refutation, are a sound inference procedure for FOP.*

**Lemma 5.7** (Lifting). *Any propositional proof produced by Naive Inference using propositional Gomory cuts can be duplicated using lifted Gomory cuts.*

Combining the two lemmas, we obtain our desired result:

**Theorem 5.8.** *Lifted Gomory cuts, together with proof by refutation, form a sound and complete inference procedure for the integer fragment of FOP.*

### 5.4 AN EXAMPLE PROOF

For a simple demonstration of a proof by lifted Gomory cuts, consider the FOP sentence $F = x(i) - 2x(S(i))$ from Sec. 3.2. An interesting fact about $F$ is that, in every satisfying model, $x(i)$ must "eventually" be zero: every time we move from $i$ to $S(i)$, the corresponding $x$ must decrease by at least a factor of two, reaching 0 in at most $\log(8) = 3$ steps. In particular, for any $i$, $x(S(S(S(i))))$ must be zero.

More formally, we can show $F \models -x(S(S(S(i))))$ using a single lifted Gomory cut: pick four copies of $F$ and one copy of the implicit clause $8 - x(i)$. Standardize apart (say, using variables $j, k, l, m, n$), and apply the substitution $\{j \to i, k \to S(i), l \to S(S(i)), m \to S(S(S(i))), n \to i\}$. Use the linear combination $(1, 2, 4, 8, 1)/16$ to get the constraint $\frac{1}{2} - x(S(S(S(i)))) \geq 0$, which leads to the cut $-x(S(S(S(i)))) \geq 0$.

## 6 EXTENSIONS

**Equality.** As defined so far, FOP cannot reason about object equality; that is, it cannot test whether two objects which appear syntactically different are in fact the same. Instead, object equality shows up only because all function and predicate expressions involving the two distinct names will evaluate to the same value. To represent equality, we could add a distinguished equality predicate $\text{equals}(x, y)$ to FOP, along with appropriate new semantics. Such a treatment would be in many ways analogous to standard extensions of FOL to include equality; for example, it would require a new inference rule, analogous to paramodulation in FOL with equality.

**Sums.** Another useful extension to FOP would be a sum quantifier: if $I$ is our context valuation, we define $\sum x. P$ as the sum, over all distinct objects $y$ from our model $M$, of the value of $P$ under $I[x \to y]$. This extension makes most sense in conjunction with object equality, since an equality predicate gives us more control over which terms appear in our sum.

Since the sum quantifier can represent a sum over infinitely many terms, we need additional restrictions to ensure convergence of the sum to a unique value. The best way to implement such restrictions is not immediately obvious, but is an interesting direction for future research.

**Inference in mixed-integer FOP.** Lifted Gomory cuts and refutation are sound and complete for integer FOP; unfortunately, the situation is more complicated for general mixed-integer sentences. Our lifted feasibility test (Sec. 5.3) works equally well for integer or mixed-integer sentences, and Lemmas 5.3 and 5.4 show that it is sound and complete in either case. However, Lemma 5.1 does not hold for mixed-integer sentences (see Gordon et al. [2009] for a counterexample), so we can no longer reduce entailment to feasibility checking by using a margin to convert from a $<$ bound to a $\leq$ bound.

To get around this problem, we can strengthen or weaken the definition of entailment:

**Definition 6.1** ($\epsilon$-entailment). *Sentence $S$ $\epsilon$-entails $S'$, $S \models_\epsilon S'$, iff* $[\text{value}(S, M) \geq 0] \Rightarrow [\text{value}(S', M) > \epsilon]$.

For $\epsilon \geq 0$, $S \models_\epsilon S'$ is stronger than $S \models S'$; for $\epsilon < 0$, it is weaker. With this altered definition, $S \wedge (-\epsilon - S')$ is feasible iff $S \models_\epsilon S'$, and so we can test $\epsilon$-entailment using our lifted feasibility procedure from Sec. 5.3. If we wish to know whether $S \models S'$, a reasonable procedure might be to test $S \models_\epsilon S'$ for one or more values of $\epsilon$; however, this procedure may give up either soundness (if we conclude entailment based on any efficiently computable $\epsilon < 0$) or completeness (if we require $\epsilon \geq 0$). On the other hand, in some applications, "approximate entailment" might be a useful conclusion in its own right.

**Direct inference of values.** Our feasibility checking procedure compares the value of a sentence to a threshold. While we can use this test together with binary search to discover tight bounds on a sentence's value, we could also ask for a procedure which discovers a sentence's value directly.

We can build such a procedure easily: given a sentence $S$, let $z$ be a new zero-argument predicate which doesn't appear in $S$. Then, the feasible region of $S' = S - z$ corresponds to the *epigraph* of $S$, that is, the set of pairs $(M, z)$ where $M$ is a model of $S$ and $z \leq \text{value}(S, M)$.

If we now run our lifted Gomory cut procedure on $S'$, each new cut will place a new bound on the feasible region of $S'$ and therefore on the epigraph of $S$. Since we are interested in the maximal feasible value of $z$, we can focus our search to try to pin down this maximal value quickly: first note that $z$ will appear in every clause of the reduced normal form of $S'$, and so will always appear in the MILP which we construct in Step 3 of the cut procedure. So, in Step 4 (where we pick a linear combination of constraints, thereby specifying a cut which removes a corner of our subproblem's relaxed feasible region), we can make sure to cut off a corner which has maximal $z$: we can find such a corner by optimizing $z$ over our relaxed feasible region, and then make a cut by choosing a row of the simplex tableau corresponding to this corner. A further benefit of this procedure is that the optimal $z$ values for each of our subproblems will be upper bounds on $\text{value}(S)$.

In applications where we care about the exact value of a sentence $S$, rather than merely its sign, we may be interested in the following extension of entailment:

**Definition 6.2** (Strong entailment). *Sentence $S$ strongly entails sentence $S'$, $S \models^* S'$, iff, for all models $M$,* $\text{value}(S', M) \geq \text{value}(S, M)$.

(We could also define $S \models^*_\epsilon S'$ to parallel Defn. 6.1.) Strong entailment implies ordinary FOP entailment, but not vice versa. If $S \models^* S'$, then the conjunction $S \wedge S'$ has the same value as $S$ in any model. By contrast, if we only have $S \models S'$, the conjunction $S \wedge S'$ only preserves whether the value is $\geq 0$ in each model.

**Concrete FOP.** One special case of FOP with equality is particularly useful in practice: if we assume *unique names* (all syntactically-distinct object constants refer to distinct objects), *known functions* (all function values are prespecified), and *domain closure* (there are no objects other than the ones mentioned in the sentence), we get a restricted version of FOP in which inference is much easier. We will call this version *concrete* FOP.

In particular, in concrete FOP, we can take the Herbrand universe for any sentence $S$ to include all and only the named object constants, and so the concrete Herbrandization of $S$ will be *finite*. There is then no need to consider finite subsets of the Herbrandized sentence: instead, we can convert the entire Herbrandized sentence to a MILP (as described in Sec. 5.2) and solve it directly. Since the MILP is finite, we can efficiently compute a threshold $\epsilon$ which allows us to reduce entailment to feasibility. And, no restrictions on the use of the $\sum$ quantifier are required, since we need not worry about convergence of sums.

In concrete FOP, quantifiers can be seen as macros: any expression containing quantifiers is equivalent to some finite, quantifier-free expression, and quantified statements merely prescribe a way to generate repetitive pieces of a ground formula. The representative power of concrete FOP is therefore no greater than that of ordinary MILPs, although equivalent problem statements may still be far more compact in concrete FOP than they are as MILPs.

Even in concrete FOP, lifted inference can still be useful, since it allows us to perform many propositional inference steps simultaneously. Many papers about "lifted inference" consider only this sort of lifting, in which we operate on a lifted representation which is more compact than the corresponding propositional representation, but still only equivalently powerful. The true power of lifted inference, however, only becomes apparent when our lifted representation is capable of representing concepts which are impossible to specify using only propositional syntax (as is true both for FOL and (non-concrete) FOP).

## 7 CONCLUSION AND FUTURE WORK

We have defined first-order programming, and established formal methods for reasoning about first-order programs, including a sound and complete inference procedure for integer FOP. Future work includes extending FOP to include equality and a summation quantifier. We also plan to experiment with lifted Gomory cuts, demonstrating their ability to make proofs more compact than corresponding proofs in FOL or MILP. Finally, we plan to implement our lifted proof-search procedure, and test it on real-world examples such as first-order stochastic programs with recourse. Finding proofs quickly in real examples will require us to generalize standard FOL search-control heuristics to FOP.

**Acknowledgements**

The authors gratefully acknowledge support from DARPA's Computer Science Study Panel program (grant HR0011-07-10026, all authors), a Lucent-Alcatel Fellowship (author SAH), and ARO grant W911NF-08-1-0301 (authors GJG and MD).

**References**

M. Alighanbari and J. P. How. Cooperative task assignment of unmanned aerial vehicles in adversarial environments. In *Proc. IEEE American Control Conference (ACC)*, 2005.

V. Borkar, V. Chandru, and S. Mitter. Mathematical programming embeddings of logic. *Journal of Automatic Reasoning*, 2002.

R. Braz, E. Amir, and D. Roth. Lifted first-order probabilistic inference. In *IJCAI-05*, 2005.

Vijay Chandru and John Hooker. *Optimization methods for logical inference*. Wiley, New York, 1999.

B. C. Eaves and U. G. Rothblum. Formulation of linear problems and solution by a universal machine. *Mathematical Programming*, 65(1–3), 1994.

R. Fourer, D. M. Gay, and B. W. Kernighan. *AMPL: A Modeling Language for Mathematical Programming*. Duxbury Press, 2002. http://ampl.com.

R. E. Gomory. Outline of an algorithm for integer solutions to linear programs. *Bull. Amer. Math. Soc.*, 64(5):275–278, 1958.

G. Gordon, S. A. Hong, and M. Dudík. First-order mixed integer linear programming. Technical Report CMU-ML-09-108, 2009.

D. Heckerman, C. Meek, and D. Koller. Probabilistic entity-relationship models, PRMs, and plate models. In Lise Getoor and Ben Taskar, editors, *Introduction to statistical relational learning*, chapter 7. MIT Press, 2007.

J. N. Hooker and M. A. Osrio. Mixed logical/linear programming. *Discrete Applied Mathematics*, 96–97:395–442, 1999.

B. Milch, B. Marthi, D. Sontag, S. Russell, and D. L. Ong. BLOG: Probabilistic models with unknown objects. In *IJCAI-05*, 2005.

A. Pfeffer. IBAL: A probabilistic rational programming language. In *IJCAI-01*, 2001.

D. Poole. The independent choice logic and beyond. *Probabilistic Inductive Logic Programming: Theory and Application, LNAI*, 2008.

W. B. Powell. A stochastic formulation of the dynamic assignment problem, with an application to truckload motor carriers. *Transportation Sci.*, 30(3):195–219, 1996.

M. Richardson and P. Domingos. Markov logic networks. *Machine Learning*, 62:107–136, 2006.

D. Roth and W. Yih. Integer linear programming inference for conditional random fields. In *ICML-05*, 2005.

S. Russell and P. Norvig. *Artificial Intelligence: A modern Approach*. Pearson Education, Inc., 2003.

T. Sandholm, A. Gilpin, and V. Conitzer. Mixed-integer programming methods for finding Nash equilibria. In *AAAI-05*, 2005.

P. Singla and P. Domingos. Markov logic in infinite domains. In *UAI-07*, 2007.

P. Singla and P. Domingos. Lifted first-order belief propagation. In *AAAI-08*, 2008.

T. Vossen, M. Ball, and R. H. Smith. On the use of integer programming models in AI planning. In *IJCAI-99*, 1999.

L. Wolsey and G. L. Nemhauser. *Integer and Combinatorial Optimization*. John Wiley & Sons, Inc., New York, 1988.